\documentclass{elsart}

\usepackage{natbib}
\usepackage{courier}

\usepackage{psfig}

\usepackage{amssymb}

\def\arcsec{\hbox{$^{\prime\prime}$}}

\begin{document}

\begin{frontmatter}

\title{The magnetic field topology associated to two M flares}

\author[IAFE]{M.L. Luoni\corauthref{cor}},
\corauth[cor]{Corresponding author, e-mail: mluoni@iafe.uba.ar}
\author[IAFE] {C.H. Mandrini\thanksref{coni}},
\thanks[coni]{Member of the Carrera del Investigador Cient\'\i fico, 
        CONICET, Argentina}
\author[IAFE]{G.D. Cristiani},
\author[Meudon]{P. D\'emoulin}
\address[IAFE] {Instituto de Astronom\'\i a y F\'\i sica del Espacio, 
        CONICET-UBA, CC. 67 Suc. 28, 
        1428 Buenos Aires, Argentina}
\address[Meudon]{Laboratoire d'Etudes Spatiales et d'Instrumentation en 
        Astrophysique (LESIA), Observatoire de Paris, 5 place Jules Janssen, 
        F-92195 Meudon Cedex, France}

\begin{abstract}
On 27 October, 2003, two GOES M-class flares occurred in the lapse of three
hours in active region NOAA 10486. The two flares were confined and their 
associated 
brightenings appeared at the same location, displaying a very similar shape
both at the chromospheric and coronal levels. 
We focus on the analysis of magnetic field (SOHO/MDI),
chromospheric (HASTA, Kanzelh{\"o}he Solar 
Observatory, TRACE) and coronal (TRACE) observations. 
By combining our data analysis with a
model of the coronal magnetic field, we compute the magnetic field topology 
associated to the two M flares.
We find that both events can be explained in terms of a localized magnetic 
reconnection process occurring at a coronal magnetic null point. This
null point is also present at the same location one day later, on 28 October, 
2003. Magnetic energy release at this null point was proposed as the origin of
a localized event that occurred 
independently with a large X17 
flare on 28 October, 2003 \citep{Mandrini06}, at 11:01~UT. The three events, 
those on 27 October and the one on 28 October, are homologous. Our results 
show that coronal null points can be stable topological structures where 
energy 
release via magnetic reconnection can happen, as proposed by classical 
magnetic reconnection models.
\end{abstract}


\end{frontmatter}

\section{Introduction}
\label{Introduction}

Activity at the solar corona, such as flares, coronal mass ejections and, 
in a low
energy release level, coronal heating, is thought to be 
related to the way in which the coronal field reacts to photospheric motions.
However, the efficiency with which the field produces energetic events 
depends on the degree of its complexity. That is to say, on the presence of 
topological structures such as null points, separators, separatrices and 
quasi-separatrix layers, which provide the small scales where magnetic 
reconnection can efficiently occur 
\citep[see the recent reviews by ][]{Demoulin05,Demoulin06,Longcope05b}.

The first models of magnetic reconnection were developed in 2D (dimensions)
\citep{Parker57,Sweet58}. In these models magnetic reconnection occurs 
at locations where the two components of the field vanished, which are
called X points. When these models are extended to 3D, keeping invariance by 
translation, the X point becomes a null line, which is structurally unstable 
(when breaking the translation invariance). 
The generalization to 3D of a 2D null point is simply a point where the three 
components of the field vanish. Contrary to null lines, these 
are in general structurally stable. Coronal null points have been found 
in few observational examples, either
associated with solar flares 
\citep[see e.g.][]{Mandrini91,Mandrini93,Gaizauskas98,Longcope98,Fletcher01b,
Longcope05a,Mandrini06} 
or with precursors to coronal mass ejections 
\citep[see e.g. ][]{Aulanier00b,Gary04}, as proposed by the breakout model for 
CMEs \citep{Antiochos99}. However, when they are searched for in a systematic 
way, they do not appear to be common features \citep{Demoulin94b}.

\begin{figure}  
\centerline{\psfig{figure=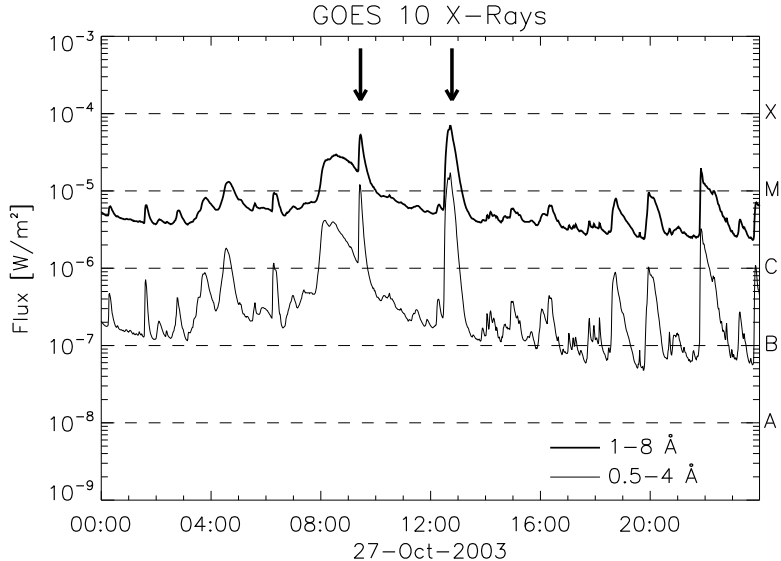,width=11.cm}}
\caption[ ]{GOES light curve for 27 October, 2003. The two analysed flares are
pointed by arrows.}    
\label{Goes}
\end{figure}

The spectacular level of activity displayed by the Sun from 19 October 
until 4 November, 2003, originated from three 
$\beta$-$\gamma$-$\delta$ sunspot groups (NOAA 10484, 10486, 10488). 
Eight of the twelve X-flares, observed during this period, 
started in active region (AR) 10486. This unusally strong activity has been
investigated in special issues of the Journal of Geophysical Research,
Geophysical Research Letters and Space Weather published in 2004 and
2005 (see: \texttt{http://www.agu.org\-/journals/ss\-/VIOLCONN1/}).
AR 10486 was also the site of other minor events (lower class) during 
that period. In particular, two M flares occurred on 27 October 
in the course of three hours. The first flare started in soft X-rays 
at 09:21 UT and was classifed as an M5.0. It appeared as an impulsive peak 
during the decay phase of a previous flare that occurred in AR 10484. The 
second flare, homologous to the first one, started at 12:27 UT  and was 
classified as an M6.7. Both events are pointed by arrows in 
Figure~\ref{Goes}.

 In this paper we concentrate in the analysis and interpretation of the two
M flares. We combine data analysis with magnetic field modelling,
which allows us to study the topology of the field at the location of the 
flares and propose a probable scenario for their origin. 
In Section~\ref{Observations}, we describe the observations; while in 
Section~\ref{Origin-flares}, we present 
our magnetic field model and topology analysis. Finally, in 
Section~\ref{Conclusions},
we show that the field topological structure is similar to the one found by 
\citet{Mandrini06} to explain the origin of the localized event that 
accompanied the two-ribbon X17 flare observed on 28 October, 2003, in the 
same AR. We discuss our
results in the context of magnetic reconnection theory. Preliminary 
results of this study have been presented elsewhere \citep{Luoni05}.

\section{Observational characteristics of the events}
\label{Observations}

\subsection{Instruments and data} 
\label{Data}

 In this paper we use full-disk level 1.5 magnetic maps
from the Michelson Doppler Imager
\citep[MDI, ][]{Scherrer95} aboard the Solar and Heliospheric
Observatory (SOHO).  These are the average of five magnetograms taken
with a cadence of 30~seconds. The maps are constructed once every 96 minutes.  
The error in the flux densities per pixel in the averaged magnetograms is 
about $\pm$ 9~G, and each pixel has a mean area of $2.08$~Mm$^2$.

 To obtain the complete temporal coverage of both flares, we use H$\alpha$ 
observations from the Kanzelh{\"o}he Solar Observatory for the flare 
in the morning, and from the H$\alpha$ Solar Telescope
for Argentina \citep[HASTA, see][]{Fernandez02} for the flare at noon on
27 October. 
Both telescopes, Kanzelh{\"o}he and HASTA, provide 
full disk images with a pixel size of 1.09\arcsec~and 
2.07\arcsec, respectively.

  Transition region and coronal observations, and also chromospheric data,
come from the Transition Region and Coronal Explorer 
\citep[TRACE, ][]{Handy99}.  TRACE was
observing with high temporal cadence (60~seconds) in the 1600~\AA\ and
195~\AA\ passbands with a FOV of 768\arcsec $\times$ 768\arcsec~at the 
times both flares occurred.  The pixel size in all images is 
0.5\arcsec, giving a spatial resolution of 1.0\arcsec.

 The plasma observed by TRACE in the 1600~\AA\ bandpass has 
temperatures in the range
of 4 - 10 $\times$ 10$^3$ K \citep[][ their Table I]{Handy99},
temperatures that correspond to the
upper photosphere and chromosphere.  As the spots are clearly visible
in these images, they have been used to coalign TRACE images with magnetic
and chromospheric data. 
 
\begin{figure}    
\centerline{\hbox{\psfig{file=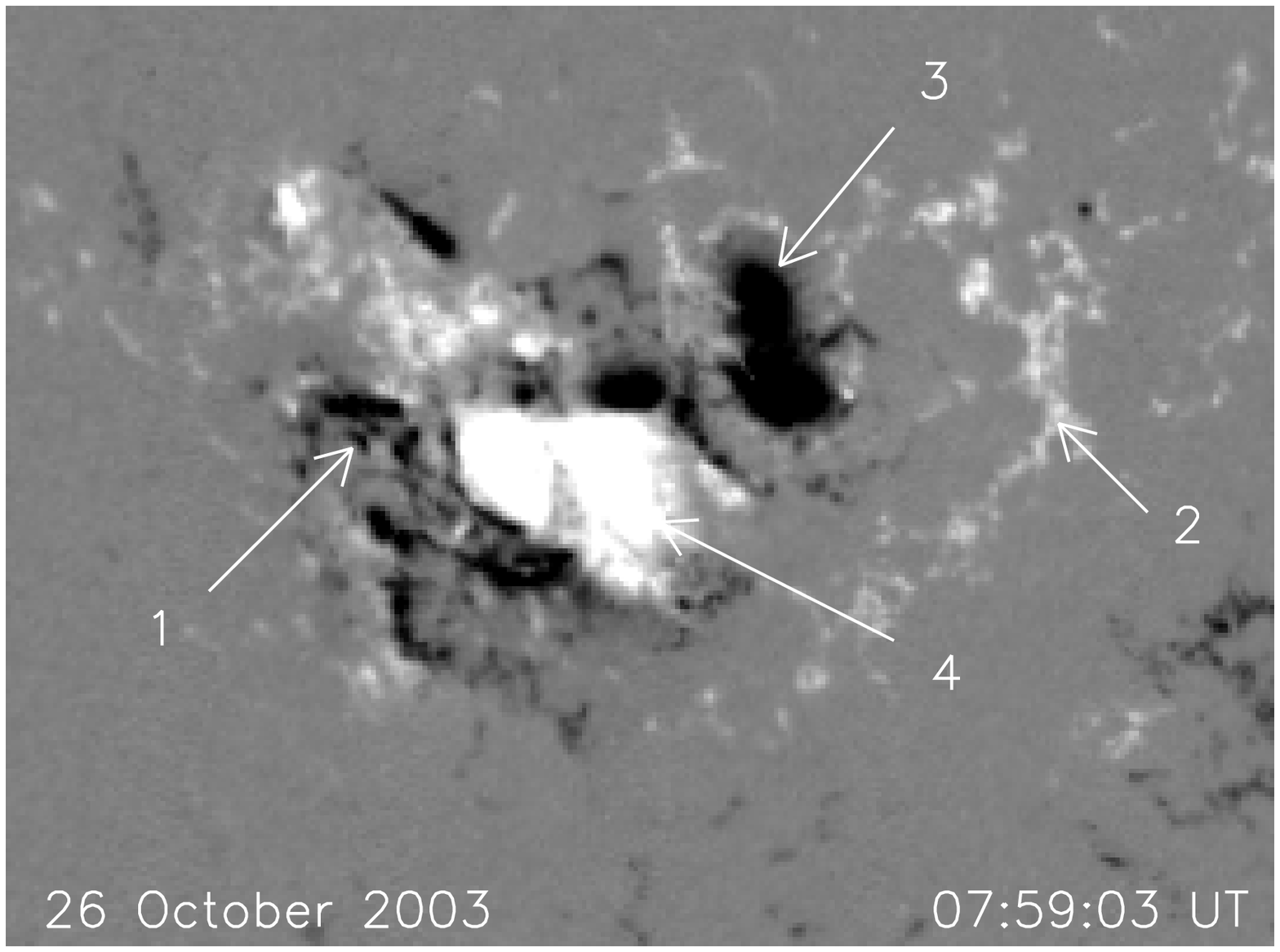,width=7.1cm,clip=}
\psfig{file=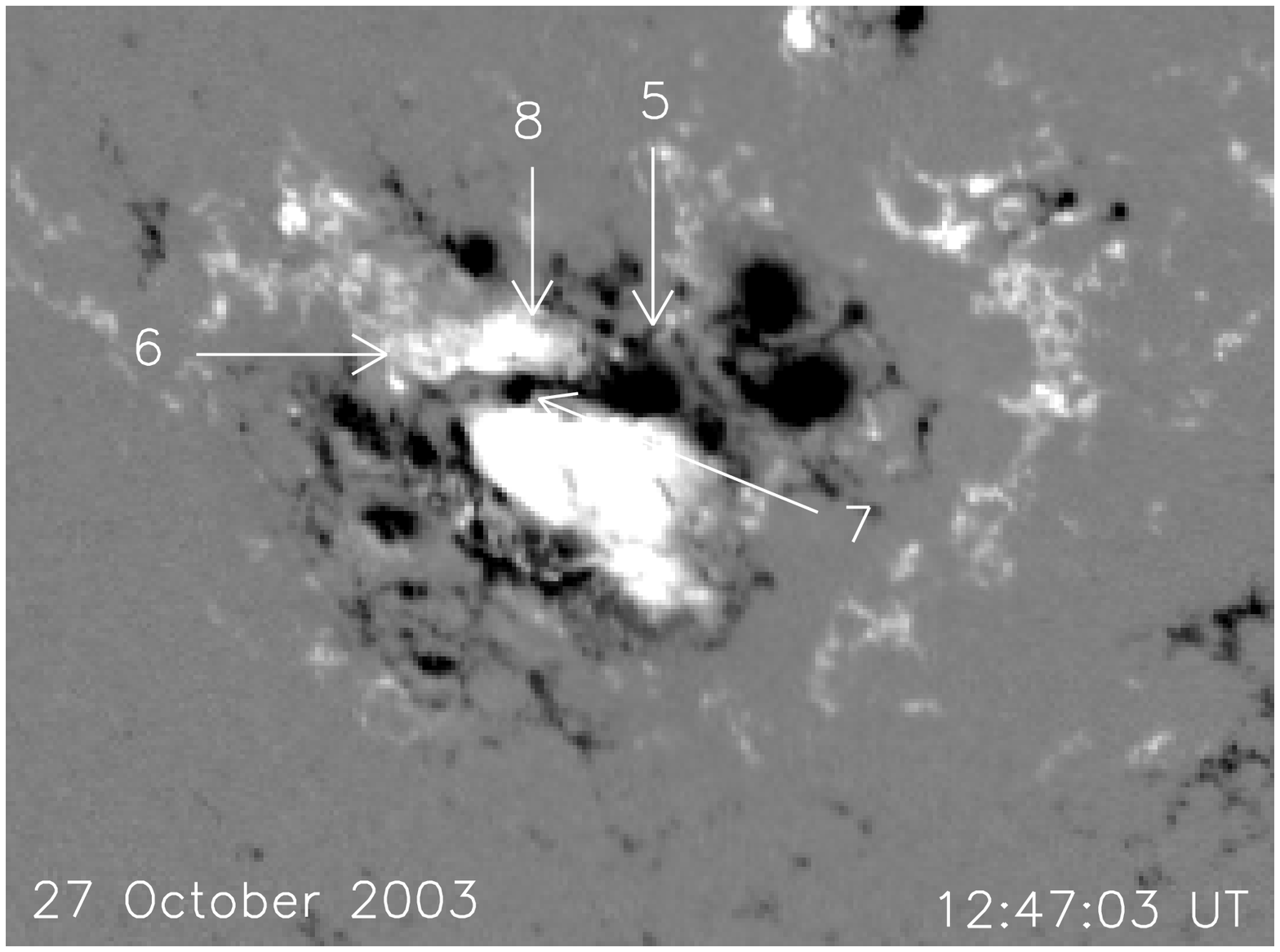,width=7.1cm,clip=}}}
\caption[]{MDI maps showing AR 10486 magnetic configuration on 26 and 27 
October, 2003. The polarities that  
are involved in the two M flares  
discussed in this paper are marked as 4, 5, 7 and 8. Other polarities
present in the AR have been numbered as described in
Section~\ref{Magnetic-obs}. The area covered by the images is 271
$\times$ 201 pixels (536\arcsec $\times$ 398\arcsec). 
The magnetic field values have been saturated
above (below) 500 G (-500 G). Positive (negative) field values are
shown in white (black) colour.  
In this and all figures depicting the
observations solar north is to the top and west is to the right.}
\label{MDI}
\end{figure}

\subsection{The magnetic field evolution} 
\label{Magnetic-obs}

   AR 10486 arrived at the East solar limb on 23 October, 2003, with
an already complex magnetic configuration.  A detailed description of the
AR magnetic field evolution 
can be found in \citet{Mandrini06}. We describe only 
what is meaningful to understand the flares discussed in this paper.
Figure~\ref{MDI} shows the AR configuration on 26 and 27 October, 2003. All
polarities in this figure have been numbered as in 
\citet[][ their Fig. 2]{Mandrini06} for comparison. 
Polarities 3 and 4 in Figure~\ref{MDI}~(left) emerged when the AR was 
still on the far side of the Sun. This emergence occurred 
within a mainly-negative field environment (to which polarity~1 belonged).  
The leading polarity~3 was located very close to the trailing 
positive polarity of an already decayed AR (polarity~2).
To the north of the positive polarity 4, several bipoles
emerged.  Polarities 5 and 6 in Figure~\ref{MDI} (right), which continued 
growing as the AR approached its central meridian passage, were already
present at the east limb. On 25 October a very small bipole
(polarities 7 and~8 in Fig.~\ref{MDI}, right; see also the small
bipole at the north of polarity 4 in the left panel)
started to emerge. By 26 October it was clearly above the main inversion 
line. This field emergence created a positive region (merging of 6 and~8) 
separated from the main positive spots (polaritiy 4) 
by an elongated negative zone (merging of 5 and~7).

\begin{figure}    
\centerline{\hbox{\psfig{file=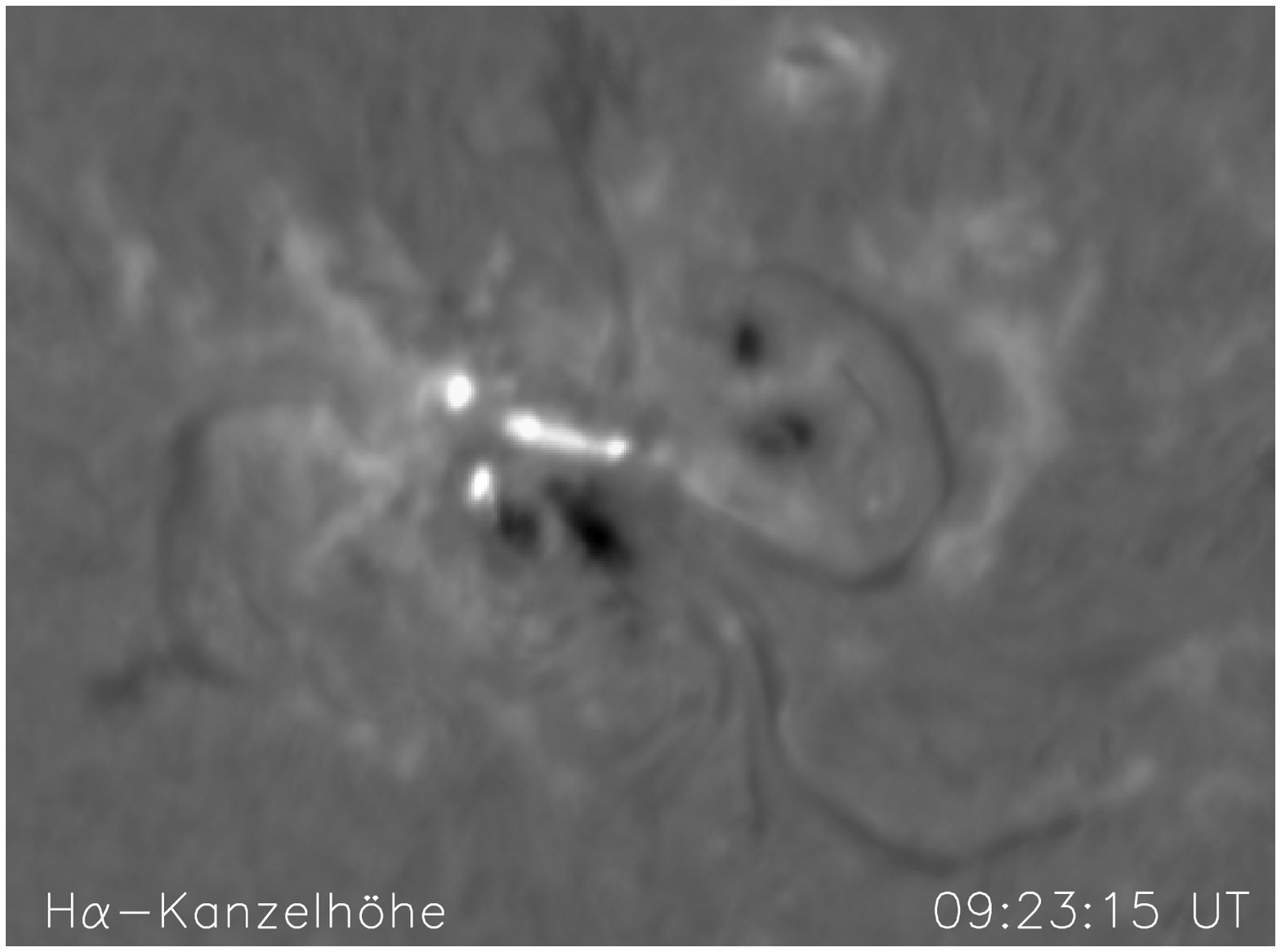,width=7.1cm,clip=}
\psfig{file=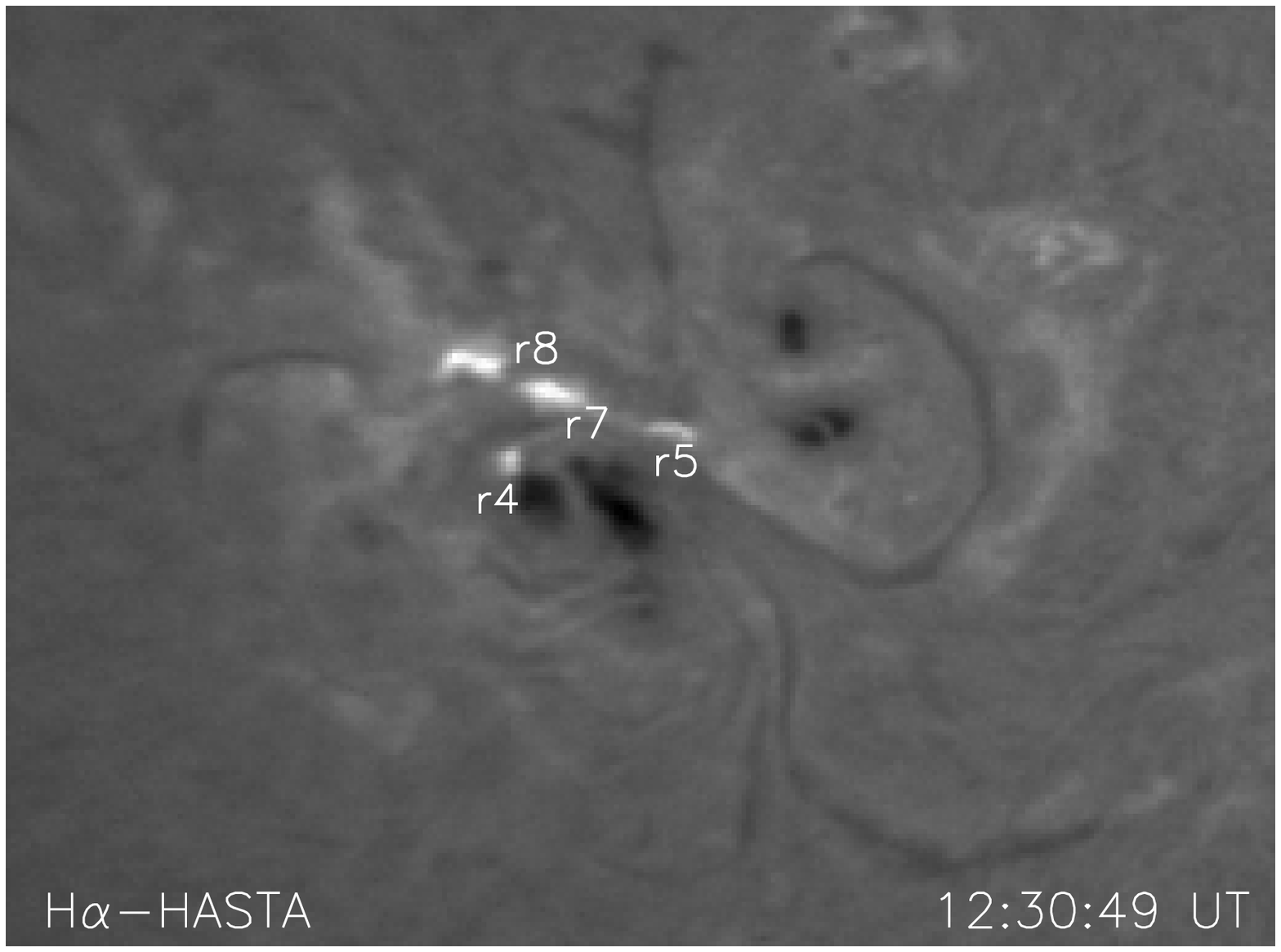,width=7.1cm,clip=}}}
\caption[]{The two M flares in H$\alpha$. 
The left panel shows an image of the first flare starting at 09:21 UT from 
Kanzelh{\"o}he Solar Observatory, while the image on the right 
is an HASTA image of the second flare starting at 12:27 UT. 
The field of view in these
images is the same as shown in Fig.~\ref{MDI}. 
The area covered by the left image 
is 491 $\times$ 364 pixels, while for the right image it is
259 $\times$ 192 pixels (536\arcsec $\times$ 398\arcsec).
Notice the similitude between 
the locations and shapes of the flare kernels. The flare ribbons have 
been marked in the right image, as discussed in the text.}
\label{Halpha}
\end{figure}

\subsection{The events at chromospheric level} 
\label{Chromospheric-obs}

\begin{figure}    
\centerline{\hbox{\psfig{file=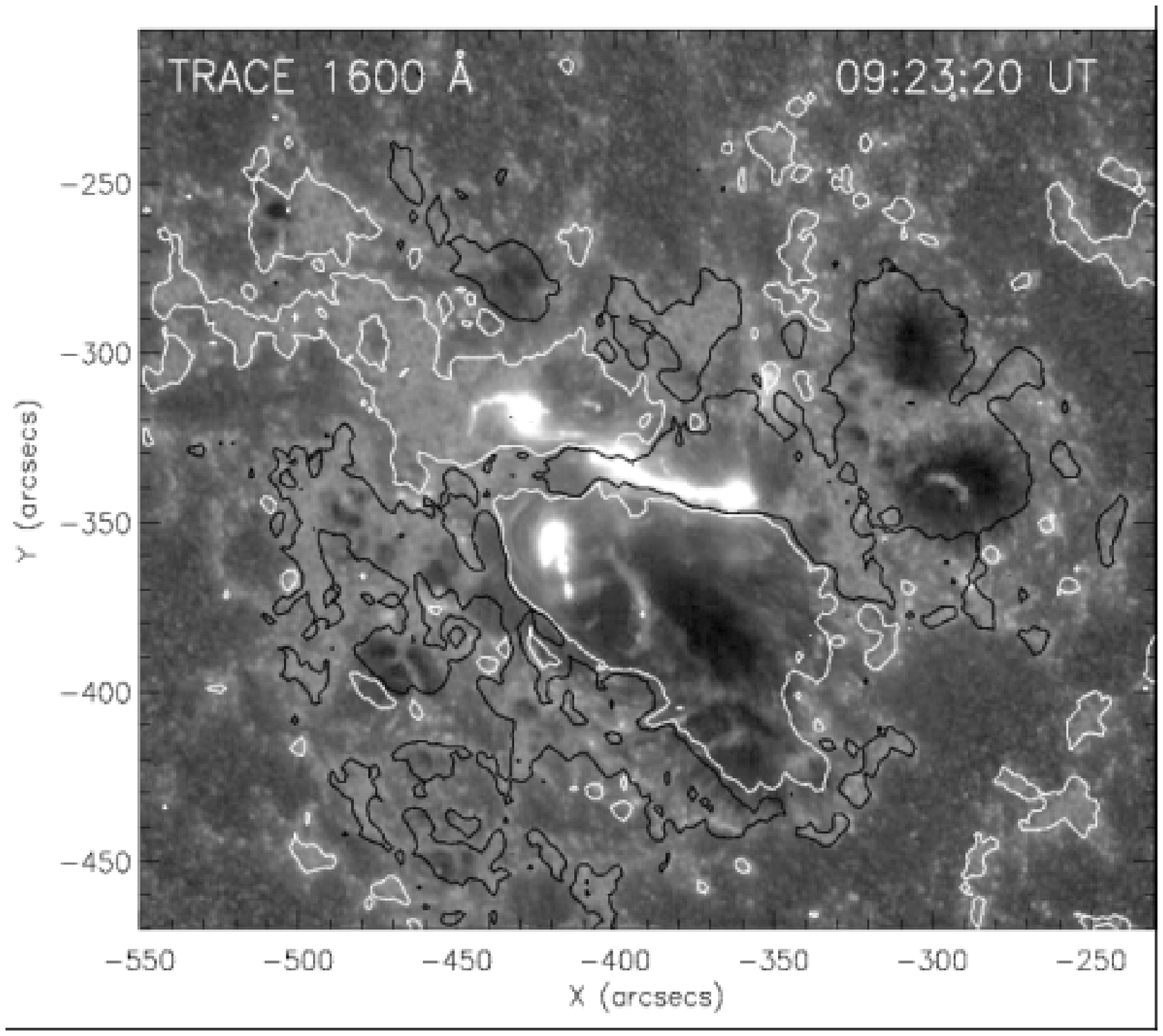,width=7.cm,clip=}
\psfig{file=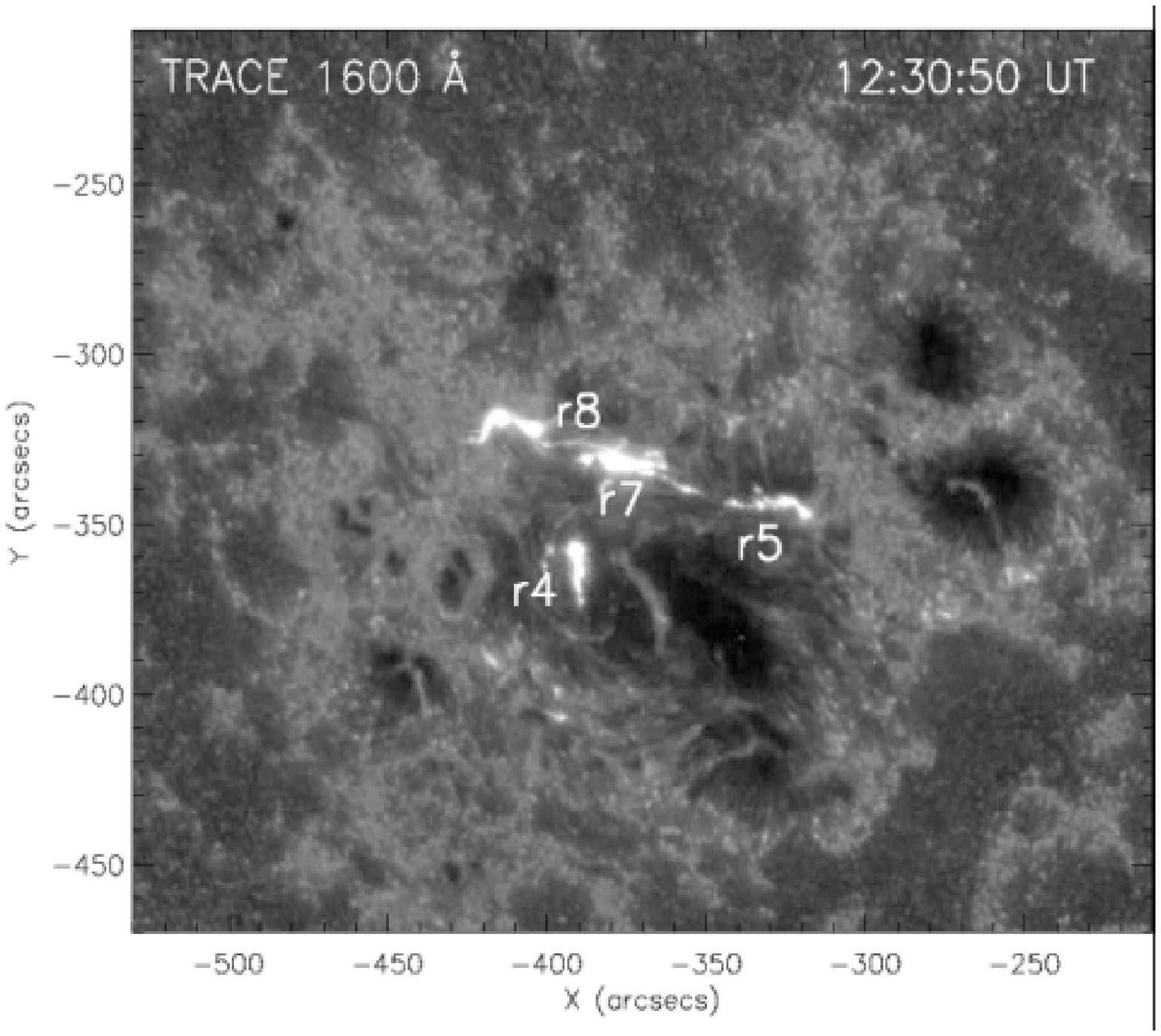,width=7.cm,clip=}}}
\caption[]{TRACE images in 1600~\AA\ showing the locations of the 
kernels of both M flares. Notice the similitude in the
shapes and locations of both flares ribbons.
An MDI (magnetogram at 9:35 UT) 
isocontour of $\pm$ 100 G (white/black continuous line corresponds 
to the positive/negative field value) has been overlaid on the image
on the left, while on the image on the right we have numbered the flare 
ribbons as discussed in the text. The field of view is the same in both
panels.}
\label{1600}
\end{figure}

  The flare during the morning of 27 October appears in Kanzelh{\"o}he 
images as four 
separated kernels or ribbons (see Fig.~\ref{Halpha}, left). 
They are located on polarities 4, 5, 7 and 8. These
brightenings appear, increase in intensity and decay at the same location 
all along the flare duration, meaning that the event is confined.
Two of these ribbons are smaller and more compact, those
on polarities 4 and 8; while the ones on 7 and 5 are more elongated and they
even merge in this case. 
In particular, the kernel on polarity 7 lies also partially on 
polarity 8; errors in the coalignment between H$\alpha$ and magnetic maps, 
that we estimate as $\pm$ 1 chromospheric image pixel,  
can be at the origin of this overlapping. The behaviour displayed by the 
H$\alpha$ observations is also seen in TRACE 1600~\AA\ images. An overlay
between a TRACE image in this band, at the same time as the Kanzelh{\"o}he 
image, and the closest in time MDI magnetic map 
can be seen in the left panel of Figure~\ref{1600}. 

  Concerning the flare on 27 October at noon, it displays an extremely similar 
behaviour as the flare in the morning, both in H$\alpha$ and 1600~\AA\, 
meaning that these two M flares are homologous.
An image from the HASTA telescope is shown in the right panel of 
Figure~\ref{Halpha}, while a TRACE image at the same time 
in 1600~\AA\ band is shown in the right panel of Figure~\ref{1600}.
 
  We have called the ribbons of both flares 
using the letter "r" for ribbon and the 
number corresponding to the polarity on which the brightening lies. This
is the same way used by \citet[][see their Figures 2 and 3]{Mandrini06} to 
call the brightenings associated with the small event that accompanied the 
two-ribbon X17 flare. During the flares on 27 October the main active region 
filament was seen to activate (its plasma was probably heated), but it 
did not erupt; its eruption occurred on the next day followed by the 
large two-ribbon flare.
   
\begin{figure}    
\centerline{\hbox{\psfig{file=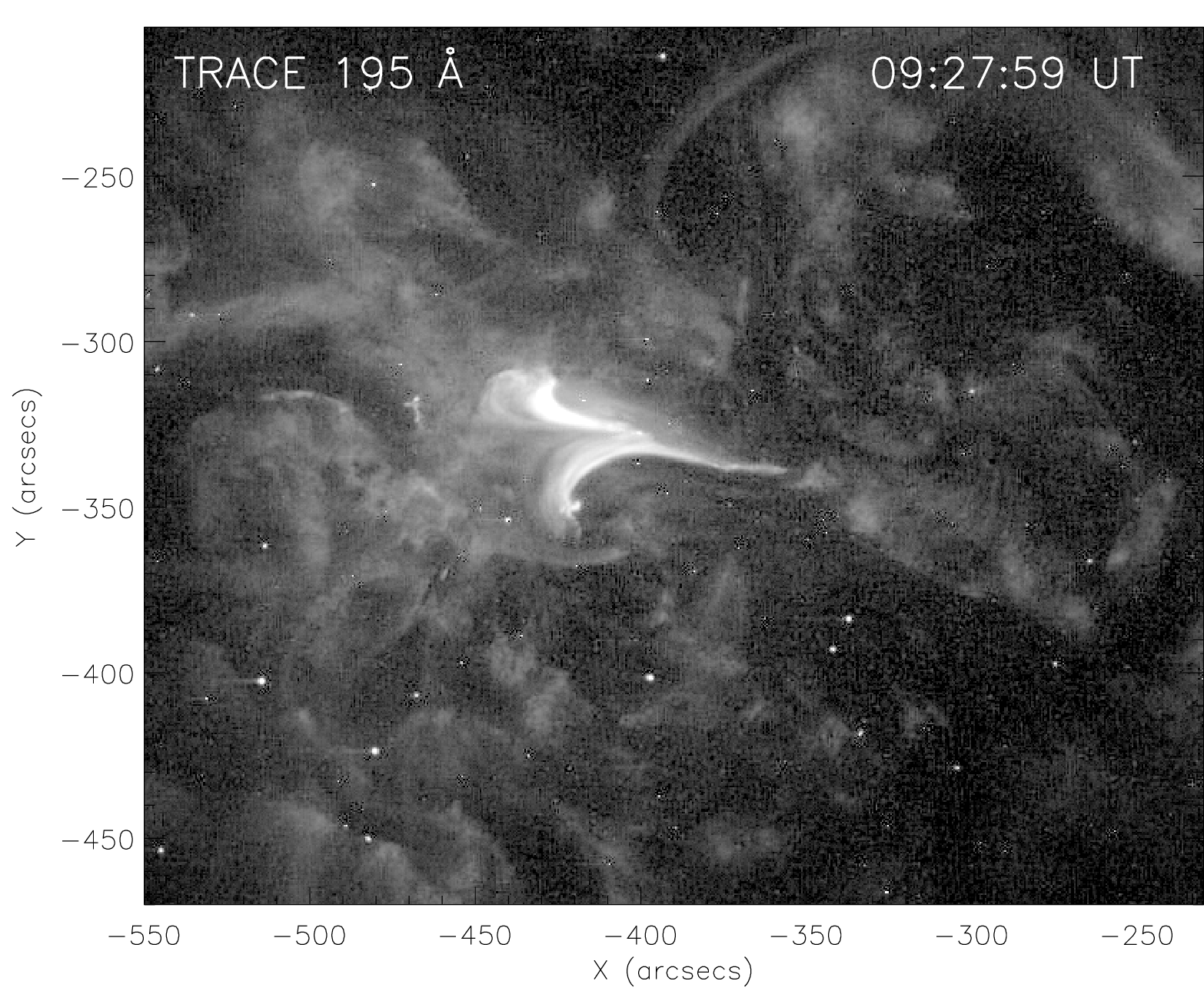,width=7.cm,clip=}
\psfig{file=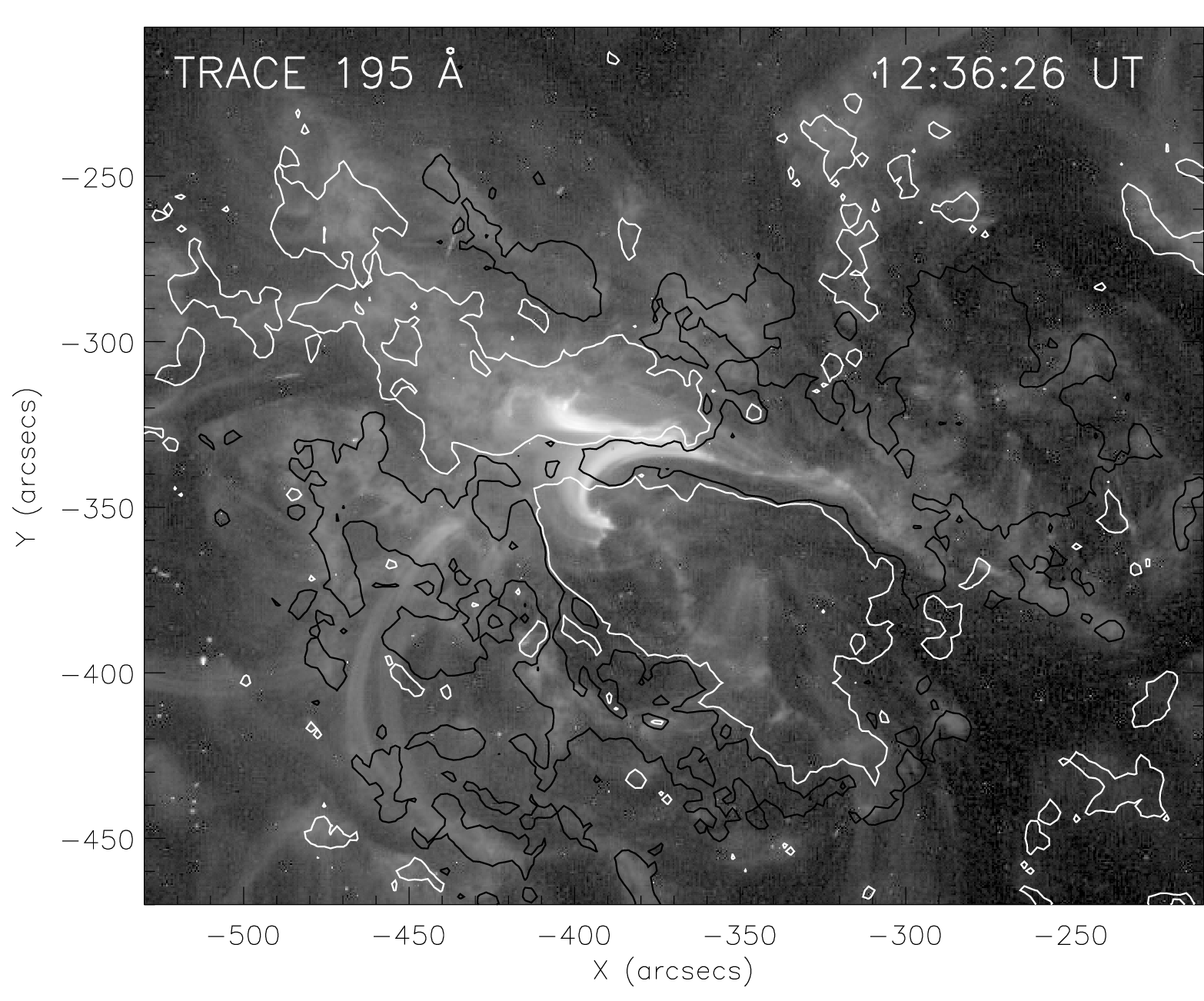,width=7.cm,clip=}}}
\caption[]{TRACE images in 195~\AA\ showing the flare 
loop brightenings at coronal level of both M flares.
Notice the similitude in the locations and shapes of both flares loops.
An MDI (magnetogram at 12:47 UT) 
isocontour of $\pm$ 100 G (white/black continuous line corresponds 
to the positive/negative field value) has been overlaid on the image
on the right as a reference. The field of view is the same in both
panels.}
\label{195}
\end{figure}

\subsection{The events at coronal level} 
\label{Coronal-obs}

  Bright coronal emission was observed by TRACE in 195~\AA\  
in the neighbourhood of polarities 7 and 8 for both M flares.  
In both cases, these brightenings had the
shape of curved loops 
and extend above the inversion lines separating
the negative elongated region from the positive polarities northward and
southward (Fig.~\ref{195}). These loops are associated
with the four ribbons (r4, r5, r7, r8) shown in Figures~\ref{Halpha}
and~\ref{1600}. This is what we expect if reconnection has 
occurred, i.e. each pair of ribbons is 
associated with a set of reconnected field lines.
These loop brightenings appear, grow in intensity and fade 
at the same location and keeping approximately the same shape 
during both flares.

\section{The origin of the homologus M flares}
\label{Origin-flares}

\subsection{The coronal-field model}  
\label{Model}

  To understand the origin of the emission described in the
previous sections, we study its relation with the 3D AR magnetic 
structure. To do so, we model the AR coronal field extrapolating 
the observed line of sight magnetic field 
under the linear force-free field assumption 
($\vec{\nabla} \times \vec{B} = \alpha \vec{B}$, with $\alpha$ constant).
We follow the method described in \citet{Demoulin97}, 
which is based in a fast Fourier transform as proposed by 
\citet{Alissandrakis81}.  This model takes
into account the transformation of coordinates from the AR location to
disk center. We have taken as boundary condition the MDI magnetogram 
in between the two M flares, the one at 11:11 UT on 27 October,
2003. 

   The only free parameter in our model is $\alpha$. We determine its
value by best fitting the loops observed by TRACE in 195~\AA\ 
(see Section~\ref{Topology} and Figs.~\ref{195} and~\ref{reconect}). 
AR 10486 magnetic field is highly non-potential
\citep{Zhang03} and, therefore, the $\alpha$ value that gives the best match
to TRACE loops 
turned out to be the largest possible allowed by our model for the
size of our integration box, $\alpha$ = -3.1 $\times$ 10$^{-2}$ Mm$^{-1}$. 
The chosen box includes all the relevant AR polarities and
is large enough to avoid aliasing effects 
\citep[see e.g.][]{Mandrini96,Demoulin97,Green02}. 

  As shown in the right panel of Figure~\ref{reconect}, our computed field 
lines follow quite well the shape of TRACE loops in Figure~\ref{195}, 
even so they do not reach as
far to the east where ribbon r8 lies. Our model is limited in the sense that 
our value of $\alpha$ is the same for all points at the photosphere. However, 
as discussed below, the same topology was found for this AR at the locations of
the flares using a non-linear force free field extrapolation 
by \citet{Regnier04}. 
Since the magnetic configuration is locally quadrupolar 
(polarities 4, 5, 7, 8), the magnetic topology is strongly defined by the 
magnetic flux distribution at the photospheric level; the magnetic field 
created by the coronal electric currents only deforms this topology.  

\subsection{The coronal-field topology}  
\label{Topology}

\begin{figure}
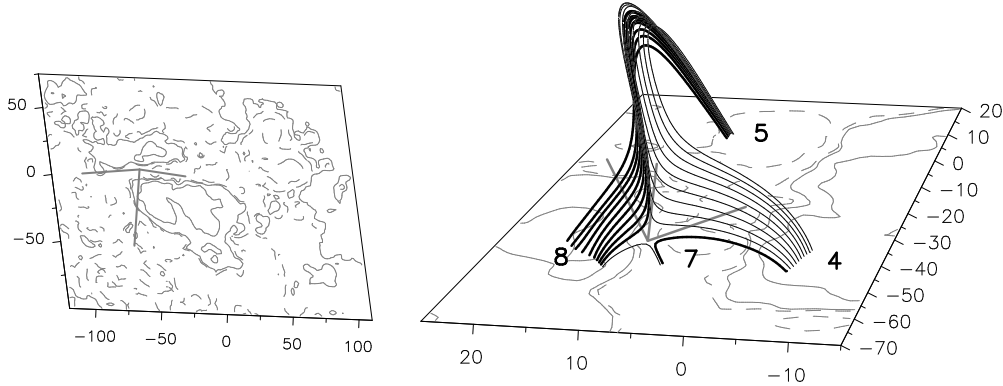
    
\centerline{\hbox{\vspace*{-0.2cm}\psfig{file=fig6a.eps,width=5.cm,clip=}
            \psfig{file=fig6b.eps,width=8.4cm,clip=}}}
\caption{The left panel shows the magnetic null point location 
in AR 10486 coronal field. This figure is in the observer's point of view. 
The right panel depicts the null location from a different point of view. 
Field lines in this panel have been computed starting integration  
at finite distances from the coronal null. 
A set of thin continuous field lines follow roughly 
the direction of the eigenvector with the lowest eigenvalue in the fan plane.
These have footpoints 
where polarities 4 and 5 lie. These field lines could reconnect at the null 
with field lines linking 7 to 8, these are represented by only one short 
and thin continuous line. After reconnection, we would have 
the set of thick continuous field lines that 
have footpoints at polarities 8 and 5, and those that connect polarities
4 and 7  (the latest are represented by only one thick short continuous field
line). The three eigenvectors of the Jacobian field matrix have been
depicted at the null location in both panels. The meaning of each of
these axes is described in the text. The negative (positive) field
isocontours are shown in continuous (dashed) thin lines, their values are 
$\pm$~100 and $\pm$~1000~G. The axes are labeled in Mm. Polarities are
indicated in both panels.
}
\label{null-obs}
\end{figure}

  We explore the coronal magnetic field configuration in search of
topological structures that can be associated to the observed flare
ribbons and loops. We find the presence of a 3D magnetic null point 
located at a height of 3.1 Mm above the magnetogram, 
over the elongated negative polarity (polarity~7). 
The location of the null is shown in the left panel of 
Figure~\ref{null-obs}. The vicinity of a null point
can be described by the linear term in the local Taylor expansion
of the magnetic field.  By diagonalizing the Jacobian matrix of the field,
we can find three orthogonal eigenvectors describing the
structure of the field in the null neighbourhood \citep{Molodenskii77}.
The divergence-free condition imposes that the sum of the three corresponding
eigenvalues vanishes ($\lambda _1 +\lambda _2 +\lambda _3 =0$).
Furthermore, if the magnetic field is 
in equilibrium with the plasma ($\vec{j} \times
\vec{B} = \vec{\nabla} P$), the eigenvalues are real \citep{Lau90}.
This means that two eigenvalues, say $\lambda _1, \lambda _2$, have the same
sign, which is opposite to that of the third eigenvalue ($\lambda _3$).  
The two field lines that
start at an infinitesimal distance from the null, in directions
parallel and anti-parallel to the eigenvector associated with $\lambda _3$, 
are called the two spines of the null. All field lines  
starting at an infinitesimal distance from the null in the plane defined by the
two eigenvectors associated with $\lambda _1$ and $\lambda _2$, define what 
is called the fan surface. 
Further information on magnetic nulls 
can be found in \citet{Greene88,Lau93,Longcope05a}.

\begin{figure}
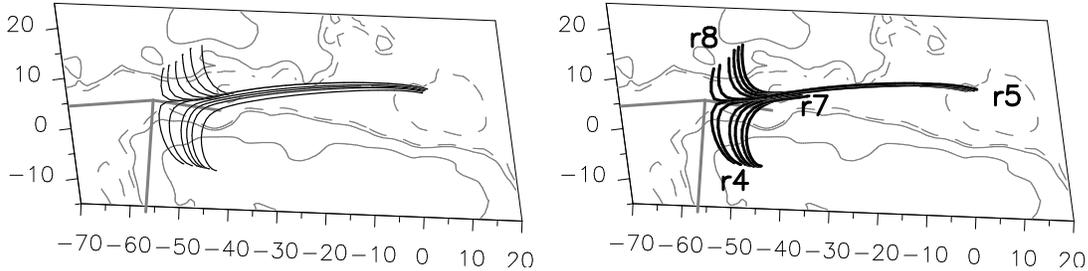
    
\centerline{\hbox{\psfig{file=fig7a.eps,width=7.2cm,clip=}
            \psfig{file=fig7b.eps,width=7.2cm,clip=}}}
\caption{Coronal magnetic field model of AR 10486 close to 
the magnetic null point. The left and 
right panels show the field lines drawn in the right panel of 
Fig.~\ref{null-obs} in the observer's point of view. 
The left panel shows the two sets of
field lines representing the pre-reconnected loops (thin continuous
lines). The right panel corresponds to
field lines
after reconnection at the null point (thick continuous lines).
Notice that the shape of these lines follows closely the shape 
of TRACE loops in 195~\AA\ in Figs.~\ref{195}.
Several short field lines have been added as compared 
to the right panel of Fig.~\ref{null-obs}. 
The isocontours of the field are 
$\pm$~100 and $\pm$~1000~G,
drawn with continuous (dashed) lines for the
positive (negative) values.  The axes are labeled in Mm.
The approximate locations of the ribbons for both flares  
are indicated in the right panel (see Fig.~\ref{Halpha} 
and Fig.~\ref{1600}), we also depict the location of the null as 
in the left panel of Fig.~\ref{null-obs}.}
\label{reconect}
\end{figure}

  The magnetic null point found in AR~10486 has two positive and
one negative eigenvalues 
(null of type B, see references in the previous paragraph).
The three axes at the location of the null in Figures~\ref{null-obs} 
and~\ref{reconect} point in the direction
of the three eigenvectors described in the previous section. The axis pointing
approximately towards the solar south corresponds to the spine of the null
(eigenvalue $\lambda _3$); while the other two axes define the 
fan plane, the one pointing approximately towards the solar west is the 
one with the lowest eigenvalue. This last eigenvelue is $\sim$ twice
lower than the other one. An study of the
field structure in the vicinity of null points created by 
theoretical configurations similar to that of AR 10486 can be found in
\citet{Mandrini06}.
In the right panel of Figure~\ref{null-obs} we have computed field
lines starting integration at finite distances from the null as described in 
the figure caption.

If magnetic reconnection occurs at the magnetic null point
shown in Figure~\ref{null-obs},
we expect to see two compact bright ribbons at the intersections
of the spine with the photosphere on the positive polarities 
4 and 8, and two additional ones,
having an elongated shape, on the negative region between them
(the photospheric trace of the fan). H$\alpha$ and TRACE
observations at 1600~\AA\ show basically this 
(see Figs.~\ref{Halpha} and~\ref{1600}).

   The above conclusion is reinforced when we draw the field lines
in the right panel of Figure~\ref{null-obs} in the
observer's point of view, as shown in Figure~\ref{reconect}.
The left panel in this
figure corresponds to field lines before
reconnection, while the ones after
reconnection are shown in the right panel.  
The shape of this last set of field lines follows
closely the shape of TRACE loops in 195~\AA\ shown in
Figures~\ref{195}.  Combining the magnetic field
evolution analysis (Section~\ref{Magnetic-obs}) with the computed
coronal field topology, we conclude that the magnetic null appears in
the corona as polarities 7 and~8 emerged and grows in the pre-existing
field of polarities 4 and~5 (Fig.~\ref{MDI}). Then, our results confirm that 
magnetic nulls are favorable regions for magnetic reconnection, which  
is driven by the local photospheric magnetic evolution.

A coronal null point was found at almost the same location on
28 October, 2003, by \citet{Mandrini06}. The null point
height above the photosphere was 4.3 Mm on that day. The   
bright short loops observed by TRACE from one hour before the large X17 flare,
and all along this event, had the same shape as the loops seen by
TRACE in 195~\AA\ in our case. Furthermore, ribbons were observed at the
same location as the flare 
ribbons observed during the M flares analysed in this paper.
The just described ribbons and loops on 28 October correspond to a 
small event that accompanied the large X17 flare.
No magnetic null point was found at this location or near by on 26 October,
2003, by \citet{Li06} who studied the magnetic field topology of AR 10486 
in relation with an X1.2 flare on that day.
Therefore, our results combined with those of \citet{Mandrini06} 
show that this null point is a stable topological structure 
which stays at least two days. Furthermore, 
\citet{Regnier04} used a non-linear force-free field
extrapolation of a magnetogram from the Imaging Vector Magnetograph 
on 27 October, and found a coronal magnetic null point at the same location.
The magnetic data and the extrapolation method are different from ours,
giving further support to the stability of this magnetic null point.

\section{Conclusion}
\label{Conclusions}

    The presence of coronal magnetic nulls implies that the
magnetic configuration has a complex topology which is 
favorable for magnetic reconnection to 
occur.  That is why, their existence has been theoretically invoked
as necessary for reconnection. However, it has been shown in many examples
that flares occur in more general topologies than those having null points, 
i.e. with quasi-separatrix layers
\citep[see the reviews by ][]{Demoulin05,Demoulin06}.

We model the coronal field of
AR~10486 using a linear force-free approach, taking
the observed longitudinal
photospheric field as boundary condition.  Since the AR field is
highly non-potential, the magnetic stress
($\alpha$ value) is set to the highest possible value for this type of
computation. Using this model,  we have found a magnetic
null at the coronal level.  It is located very low down in the
corona, at a height of  
$\approx 3$~Mm above the photosphere.
Computing field lines starting at finite distances from the
null, we
are able to reproduce the special shapes for two sets of observed
coronal loops for two M flares that occurred on 27 October, 2007.  
We conclude that these small loops were formed
by magnetic reconnection during both homologous M flare.  
The correspondence between the computed
field lines, located in the vicinity of a magnetic null, with the
observed loops also validates our magnetic extrapolation. 

A null point
was also present at the same location one day later, on 28 October, 2003.
Magnetic reconnection at this null point was at the origin of the small 
confined event that accompanied the large two-ribbon flare on that day 
\citep{Mandrini06}. The existence of a null point at almost the same 
location, though at different coronal heights (a change logically 
given by the photospheric evolution), demonstrates that 
this coronal null 
points is a stable topological structure. The origin of this stability is 
the presence of a quadrupolar region (polarities 4, 5, 7, 8) which globally 
stays unchanged and defines the coronal magnetic topology. The modification 
of the magnetic shear simply deforms this magnetic topology 
shifting slightly the location of the magnetic null and its separatrices. 

{\bf Acknowledgments:}
  The authors thank the Kanzelh{\"o}he Solar Observatory for providing 
the H$\alpha$ observations for one of the flares, 
and the SOHO/MDI and TRACE consortia for their data. SOHO is a project of 
international cooperation between ESA and NASA. 
This study is partially based on data obtained at OAFA (El Leoncito, 
San Juan, Argentina) in the framework of the German-Argentinean HASTA/MICA 
Project, a collaboration of MPE, IAFE, OAFA and MPAe. 
C.H.M. and  P.D. acknowledge financial support from CNRS (France)
and CONICET (Argentina) through their cooperative science program (05ARG0011
N$^0$ 18302). C.H.M., M.L.L. and G.D.C. thank 
the Argentinean grants: UBACyT X329 (UBA), PICT 12187 (ANPCyT) and 
PIP 6220 (CONICET). G.D.C. is a fellow of ANPCyT.

\bibliographystyle{elsart-harv}  
\bibliography{luoni2007a}  

\end{document}